\documentclass[amssymb,aps,preprint,prb]{revtex4}
\usepackage{graphicx}
\usepackage{graphicx,amsfonts}
\usepackage{bm}

\begin{document}

\title{Probing the superconducting gap symmetry of PrRu$_{4}$Sb$_{12}$: A
comparison with PrOs$_{4}$Sb$_{12}$}
\author{Elbert E. M. Chia}
\author{M. B. Salamon}
\affiliation{Department of Physics, University of Illinois at Urbana-Champaign, 1110 W.
Green St., Urbana IL 61801}
\author{H. Sugawara}
\author{H. Sato}
\affiliation{Department of Physics, Tokyo Metropolitan University, Hachioji, Tokyo
192-0397, Japan}
\date{\today}

\begin{abstract}
We report measurements of the magnetic penetration depth $\lambda$
in single crystals of PrRu$_{4}$Sb$_{12}$ down to 0.1~K. Both
$\lambda$ and superfluid density $\rho_{s}$ exhibit an exponential
behavior for $T$ $<$ 0.5$T_{c}$, with parameters
$\Delta$(0)/\textit{k}$_{B}$\textit{T}$_{c}$ = 1.9 and
$\lambda(0)$ = 2900 \AA. The value of $\Delta$(0) is consistent
with the specific-heat jump value of $\Delta C/\gamma T_{c}$ =
1.87 measured elsewhere, while the value of $\lambda(0)$ is
consistent with the measured value of the electronic heat-capacity
coefficient $\gamma$. Our data are consistent with
PrRu$_{4}$Sb$_{12}$ being a moderate-coupling, fully-gapped
superconductor. We suggest experiments to study how the nature of
the superconducting state evolves with increasing Ru substitution
for Os.
\end{abstract}

\maketitle

The recent discovery \cite{Bauer02,Maple02} of the Heavy Fermion
(HF) skutterudite superconductor (SC) PrOs$_{4}$Sb$_{12}$ has
attracted much interest due to its differences with the other
HFSC. Measurements of dc magnetic susceptibility, specific heat,
electrical resistivity and inelastic neutron scattering showed
that the ninefold degenerate $J=4$ Hund's rule multiplet of Pr is
split by the cubic crystal electric field, such that its ground
state is a \textit{nonmagnetic} $\Gamma_{3}$ doublet, separated
from the first excited state $\Gamma_{5}$ by $\sim$ 10~K. Hence
its HF behavior, and consequently the origin of its
superconductivity, might be attributed to the interaction between
the electric quadrupolar moments of Pr$^{3+}$ and the conduction
electrons. It is thus a candidate for the first superconductor
mediated by quadrupolar fluctuations, i.e. by neither
electron-phonon nor, as with other HFSC, magnetic interactions.

Surprisingly, replacement of Os by Ru, i.e. in
PrRu$_{4}$Sb$_{12}$, yields a superconductor with $T_{c} \approx
1.25$~K with significantly different properties. From the slope of
the upper critical field \cite{Takeda00} near $T_{c}$,
(--d$H_{c2}$/d$T$)$_{T_{c}}$ = 2.4 kOe/K and using the procedure
in Ref.~\onlinecite{Bauer02}, we get the effective mass of the
heavy electrons $m^{\ast} \approx 20m_{e}$. This contrasts with
the value $m^{\ast}$/$m_{e}$=45 for PrOs$_{4}$Sb$_{12}$, showing
that while PrOs$_{4}$Sb$_{12}$ is clearly a heavy-fermion
material, PrRu$_{4}$Sb$_{12}$ is at most, a marginal HF. Various
experimental results suggest that these two materials have
different order-parameter symmetry. Firstly, there is an absence
of a Hebel-Slichter peak in the nuclear quadrupole resonance (NQR)
data \cite{Kotegawa03} for PrOs$_{4}$Sb$_{12}$, while a distinct
coherence peak was seen \cite{Yogi03} in the Sb-NQR 1/$T_{1}$ data
for PrRu$_{4}$Sb$_{12}$. Secondly, the low-temperature power-law
behavior seen in specific heat \cite{Bauer02} and penetration
depth \cite{Chia03b}, and the angular variation of thermal
conductivity \cite{Izawa03}, suggest the presence of nodes in the
order parameter of PrOs$_{4}$Sb$_{12}$. Specifically,
Refs.~\onlinecite{Chia03b} and \onlinecite{Izawa03} reveal the
presence of \textit{point} nodes on the Fermi surface (FS). For
PrRu$_{4}$Sb$_{12}$, however, an exponential decrease in 1/$T_{1}$
is observed below the Hebel-Slichter peak \cite{Yogi03} and was
fit with an isotropic gap of magnitude $\Delta
(0)$=1.5$k_{B}T_{c}$, where $\Delta (0)$ is the magnitude of the
zero-temperature superconducting gap. Thirdly, muon spin rotation
($\mu$SR) experiments on PrOs$_{4}$Sb$_{12}$ reveal the
spontaneous appearance of static internal magnetic fields below
\textit{T}$_{c}$, providing evidence that the superconducting
state is a time-reversal-symmetry-breaking (TRSB) state
\cite{Aoki03}. Such experiments have not been performed on
PrRu$_{4}$Sb$_{12}$.

In this paper, we present high-precision measurements of penetration depth $%
\lambda $(\textit{T}) of PrRu$_{4}$Sb$_{12}$ at temperatures down
to 0.1~K using the same experimental conditions \cite{Chia03b} as
for PrOs$_{4}$Sb$_{12}$. Both $\lambda $(\textit{T}) and
superfluid density $ \rho_{s}$(\textit{T}) exhibit exponential
behavior at low temperatures, suggesting the presence of an
isotropic superconducting gap on the FS. Data are best fit by the
parameters $\Delta$(0)/\textit{k}$_{B}$\textit{T}$_{c}$ = 1.9 and
$\lambda(0)$ = 2900 \AA, and thus suggest that PrRu$_{4}$Sb$%
_{12}$ is a moderate-coupling, fully-gapped superconductor. The
values of $\Delta$(0)/\textit{k}$_{B}$\textit{T}$_{c}$ and
$\lambda(0)$ are consistent with values \cite{Takeda00} derived
from the specific-heat jump $\Delta C/\gamma T_{c}$=1.87, and the
linear specific-heat coefficient $\gamma$.

Details of sample growth and characterization are described in Ref.~%
\onlinecite{Takeda00}. The observation of the de Haas-van Alphen
(dHvA) effect from the same batch of samples \cite{Matsuda02}, and
the large residual resistivity ratio (RRR $\sim$ 76), reflect the
high quality of the samples. Measurements were performed utilizing
a 21-MHz tunnel diode oscillator \cite{Bonalde2000} with a noise
level of 2 parts in 10$^{9}$ and low drift. The magnitude of the
ac field was estimated to be less than 40~mOe. The cryostat was
surrounded by a bilayer Mumetal shield that reduced the dc field
to less than 1~mOe. The sample was mounted, using a small amount
of GE varnish, on a single crystal sapphire rod. The other end of
the rod was thermally connected to the mixing chamber of an Oxford
Kelvinox 25 dilution refrigerator. The sample temperature is
monitored using a calibrated RuO$_{2}$ resistor at low
temperatures (\textit{T}$_{base}$ to 1.8~K).

The deviation $\Delta \lambda $(\textit{T}) = $\lambda
$(\textit{T}) -- $\lambda $(0.1~K) is proportional to the change
in resonant frequency $\Delta $\textit{f}(\textit{T}) of the
oscillator, with the proportionality factor \textit{G} dependent
on sample and coil geometries. We determine \textit{G} for a pure
Al single crystal by fitting the Al data to extreme nonlocal
expressions and then adjust for relative sample dimensions
\cite{Chia03}. Testing this approach on a single crystal of Pb, we
found good agreement with conventional BCS expressions. The value
of \textit{G} obtained this way has an uncertainty of $\pm$10\%
because our sample, with approximate dimensions 0.7 $\times$ 0.5
$\times$ 0.25 mm$^{3}$, has a rectangular, rather than square,
basal area \cite{Prozorov2000}.

\begin{figure} \centering \includegraphics[width=16cm,clip]{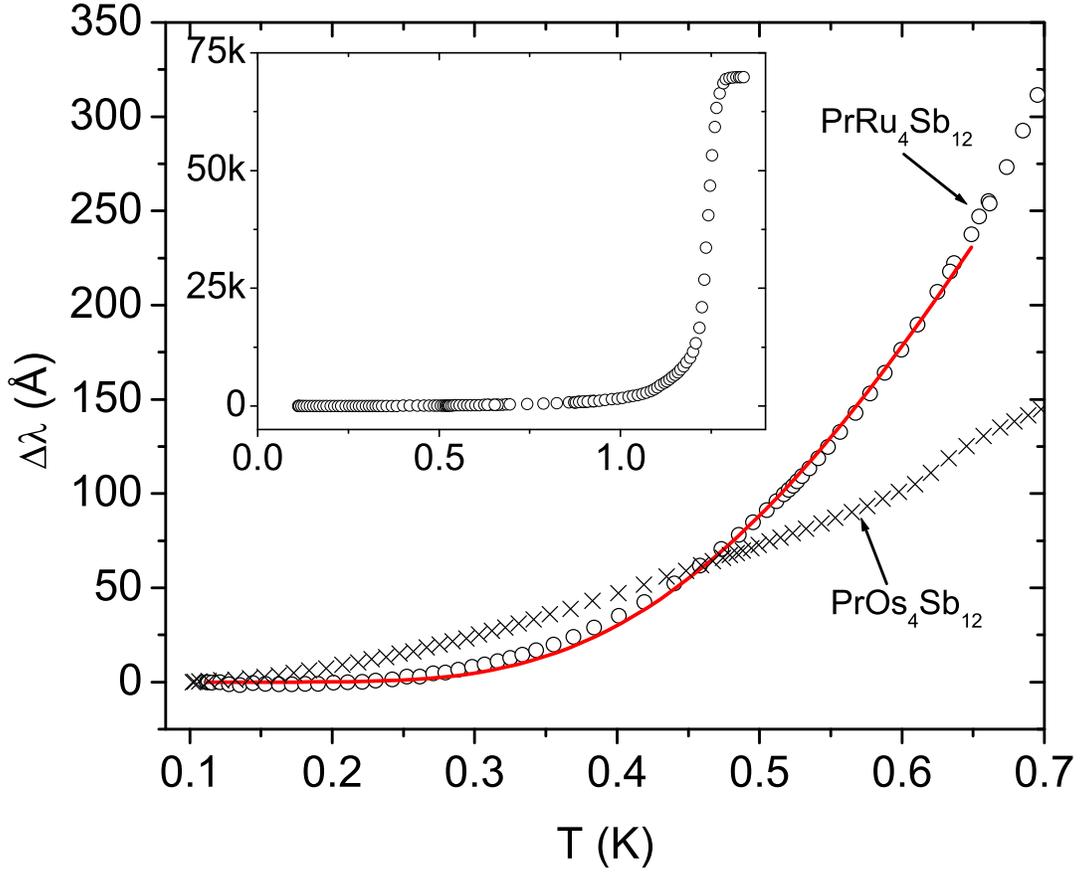}
\caption{($\bigcirc$) Low-temperature dependence of the
penetration depth $\Delta \lambda$(\textit{T}) of
PrRu$_{4}$Sb$_{12}$. The solid line is the fit to
Eqn.~\ref{eqn:lowTexponential} from 0.1~K to 0.65~K, with $\Delta
(0)$ = 2.4~K = 1.9$k_{B}T_{c}$. Inset shows $\Delta
\lambda$(\textit{T}) of PrRu$_{4}$Sb$_{12}$ over the full
temperature range. ($\times$) $\Delta \lambda$(\textit{T}) data of
PrOs$_{4}$Sb$_{12}$, taken from Ref.~\onlinecite{Chia03b}.}
\label{fig:lambda}
\end{figure}

Figure~\ref{fig:lambda} ($\bigcirc$) shows $\Delta \lambda
$(\textit{T}) for PrRu$_{4}$Sb$_{12}$ as a function of temperature
in the low-temperature region. The inset shows $\Delta \lambda
$(\textit{T}) for the entire temperature range. The value of
$T_{c}$, taken to be the mid-point of the transition, is 1.25~K.
The data points flatten out below 0.22~K ($\sim 0.18T_{c}$),
implying activated behavior in this temperature range. Contrasting
this, we superpose data for PrOs$_{4}$Sb$_{12}$ on the same figure
($\times$), which show $\Delta \lambda$ varying strongly ($\sim
T^{2}$) with temperature, indicative of low-lying excitations
\cite{Chia03b}. We fit the PrRu$_{4}$Sb$_{12}$ data to the BCS
low-temperature expression in the clean and local limit, from
$T_{base}$ (0.1~K $\approx 0.08~T_{c}$) to 0.65~K ($\approx 0.5
T_{c}$), using the expression
\begin{equation} \label{eqn:lowTexponential}
\Delta \lambda (T) \propto \sqrt{\frac{\pi \Delta
(0)}{2k_{B}T}}exp\left(-\frac{\Delta (0)}{k_{B}T}\right),
\end{equation}
with the proportionality constant and $\Delta (0)$ as parameters.
The best fit (solid line in Fig.~\ref{fig:lambda}) is obtained
when $\Delta (0)$ = 2.4~K = 1.9$k_{B}T_{c}$. This value is larger
than the BCS weak-coupling value of 1.76$k_{B}T_{c}$, suggesting
that PrRu$_{4}$Sb$_{12}$ is in the moderate-coupling regime. To
check the validity of this value of $\Delta (0)$ we make use of
the strong-coupling equations \cite{Orlando1979, Kresin1975}
\begin{equation}
\eta _{\Delta }(\omega _{0})=1+5.3\left(\frac{T_{c}}{\omega _{0}}\right)^{2}\ln \left(\frac{%
\omega _{0}}{T_{c}}\right), \label{eqn:strongcouplingDelta}
\end{equation}

\begin{equation}
\eta _{Cv}(\omega _{0})=1+1.8\left(\frac{\pi T_{c}}{\omega _{0}}\right)^{2}\left(\ln (\frac{%
\omega _{0}}{T_{c}})+0.5\right), \label{eqn:strongcouplingCv}
\end{equation}

\begin{equation}
\eta _{\lambda }(\omega _{0})=\frac{\sqrt{1+(\frac{\pi T_{c}}{\omega _{0}}%
)^{2}(0.6\ln (\frac{\omega _{0}}{T_{c}})-0.26)}}{1+(\frac{\pi
T_{c}}{\omega _{0}})^{2}(1.1\ln (\frac{\omega
_{0}}{T_{c}})+0.14)}, \label{eqn:strongcouplinglambda}
\end{equation}
where each $\eta $ represents the correction factor to the
corresponding weak-coupling BCS value. If we take $\Delta
(0)$=1.9$k_{B}T_{c}$, then Eq.~(\ref{eqn:strongcouplingDelta})
gives the characteristic (equivalent Einstein) frequency
$\omega_{0} \approx$ 17~K and Eq.~(\ref{eqn:strongcouplingCv})
gives $\Delta C/\gamma T_{c}$=1.9. This value of $\Delta C/\gamma
T_{c}$ agrees excellently with the measured value (1.87) in
Ref.~\onlinecite{Takeda00}, giving further evidence that
PrRu$_{4}$Sb$_{12}$ is indeed a moderate-coupling superconductor.

To extract the superfluid density $\rho_{s}$ from our data, we
need to know $\lambda (0)$. For a type-II superconductor, $\lambda
(0)$ can be obtained from \cite{Gross1986}
\begin{equation} \label{eqn:lambda0}
\lambda (0) = \frac{[\Phi_{0}H_{c2}(0)]^{1/2}}
{\sqrt{24}\delta_{sc}T_{c}\gamma^{1/2}},
\end{equation}
where $\Phi_{0}$=2.06$\times$10$^{9}$~G$\cdot$\AA$^{2}$ is the
flux quantum, $H_{c2}(0)$ is the upper critical field at $T$=0,
$\delta_{sc}$$\equiv$$\Delta (0)/k_{B}T_{c}$, and $\gamma$ is the
electronic specific heat coefficient. Using
$(-dH_{c2}/dT)_{T_{c}}$=2.4~kOe/K from Ref.~\onlinecite{Takeda00}
and the expression
$H_{c2}(0)$=0.693($-$d$H_{c2}$/d$T$)$_{T_{c}}T_{c}$, we obtain
$H_{c2}(0)$=2.16~kOe. The superconducting coherence length
$\xi_{0}$ can be estimated from the relation
$H_{c2}(0)$=$\Phi_{0}/2\pi \xi_{0}^{2}$, yielding $\xi_{0}
\approx$ 400 \AA. The value $\gamma $=59~mJ/mol$\cdot$K$^{2}$ is
also obtained from Ref.~\onlinecite{Takeda00}. We calculate
$\lambda (0)$ using two methods: (1) Taking $\delta_{sc}$=1.9, or
(2) taking $\delta_{sc}$=1.76 along with strong-coupling
corrections (Eqs.~\ref{eqn:strongcouplingDelta} and
\ref{eqn:strongcouplinglambda}). Both methods yield $\lambda (0)
\approx$ 3200 \AA. This puts PrRu$_{4}$Sb$_{12}$ in the local
limit. Furthermore, from dHvA data \cite{Matsuda02}, we estimate
the mean free path $l$$\approx$1300~\AA, implying that the sample
is close to the clean limit. To calculate $\rho_{s}$ for an
isotropic $s$-wave superconductor in the clean and local limits we
use the expression
\begin{equation} \label{eqn:rhosxi}
\rho_{s} = 1 + 2\int^{\infty}_{0} \frac{\partial f}{\partial E}
d\varepsilon,
\end{equation}
where $f$ = [exp(E/{\it k}$_{B}T$)+1]$^{-1}$ is the Fermi
function, and $E = [\varepsilon^{2}$ + $\Delta (T)^{2}$]$^{1/2}$
is the quasiparticle energy. The temperature-dependence of $\Delta
(T)$ can be obtained by using \cite{Gross1986}
\begin{equation} \label{eqn:gapinterpolate}
\Delta(\mathit{T})=\delta
_{sc}\mathit{kT}_{c}\tanh\left\{\frac{\pi}{\delta
_{sc}}\sqrt{a\left(\frac{\Delta C}{C}\right)
\left(\frac{T_{c}}{T}-1\right)}\right\},
\end{equation}
where $\delta _{sc}$ is the only variable parameter, $T_{c}$ =
1.25~K, \textit{a} = 2/3, and the specific heat jump $\Delta C/C
\equiv \Delta C/\gamma T_{c}$~=~1.87 is an experimentally obtained
value \cite{Takeda00}.

\begin{figure}
\centering \includegraphics[width=16cm,clip]{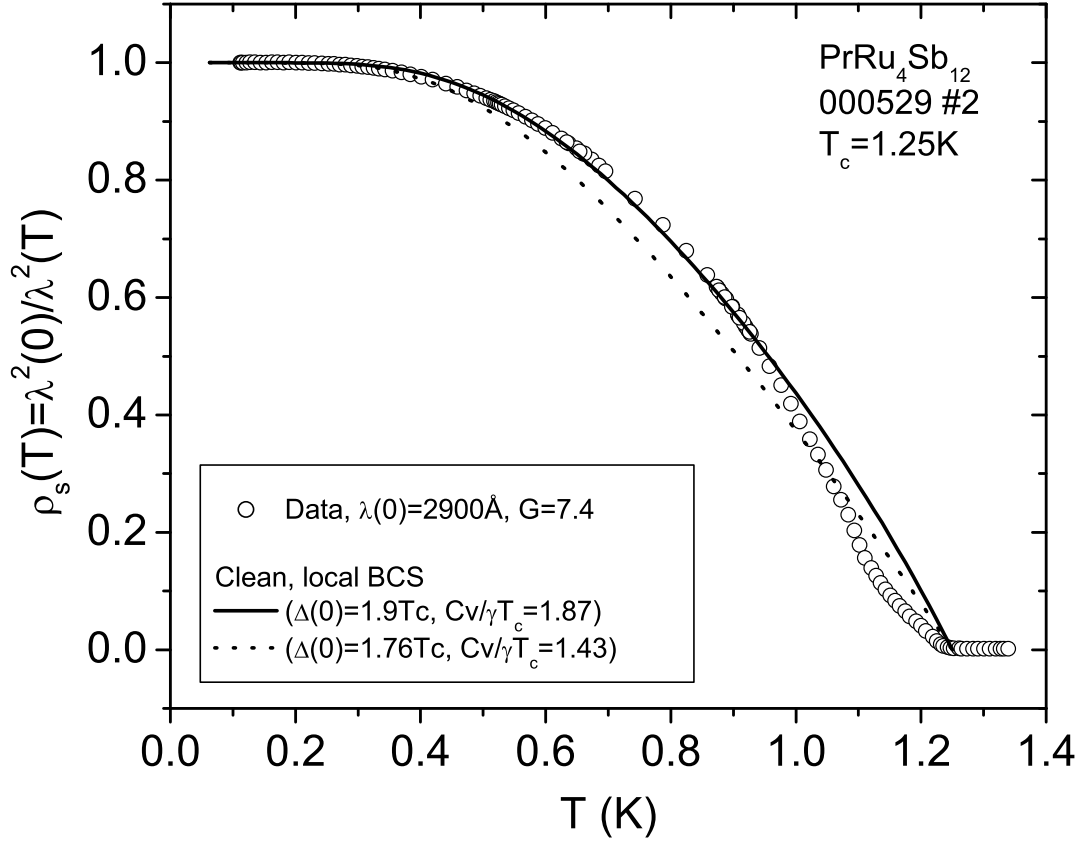}
\caption{($\bigcirc$) Superfluid density $\rho_{s}(T)$ =
[$\lambda^{2}$(0)/$\lambda^{2}(T)$] calculated from $\Delta
\lambda (T)$ data in Fig.~\ref{fig:lambda}. Lines: $\rho_{s}(T)$
calculated from Eqn.~\ref{eqn:rhosxi} with parameters $\Delta
(0)/k_{B}T_{c} = 1.9$ and $\Delta C/\gamma T_{c} = 1.87$ (solid
line), $\Delta (0)/k_{B}T_{c} = 1.76$ and $\Delta C/\gamma T_{c} =
1.43$ (dotted line).} \label{fig:rho}
\end{figure}

Fig.~\ref{fig:rho} shows the experimental ($\bigcirc$) and
calculated (solid and dotted lines) values of $\rho_{s}$ as a
function of temperature. The best fit from 0.1~K to 0.95~K
($\sim$0.8$T_{c}$) is obtained when $\lambda (0)$=2900~\AA\ and
$\Delta (0)$=1.9$k_{B}T_{c}$ (solid line). This value of $\lambda
(0)$ is 10\% lower than the earlier-calculated value of 3200~\AA,
but is acceptable because of the uncertainty in obtaining the
calibration factor $G$. The value of $\Delta (0)$ once again
agrees with the specific-heat jump $\Delta C/\gamma T_{c}$
obtained in Ref.~\onlinecite{Takeda00} via
Eqs.~(\ref{eqn:strongcouplingDelta}) and
(\ref{eqn:strongcouplingCv}), though it disagrees with the
weak-coupling value of 1.5$k_{B}T_{c}$ in
Ref.~\onlinecite{Yogi03}. The dotted line in Fig.~\ref{fig:rho},
calculated using the weak-coupling parameters $\Delta
(0)$=1.76$k_{B}T_{c}$ and $\Delta C/\gamma T_{c}$=1.43, clearly
does not fit the data. Our superfluid data once again suggest that
PrRu$_{4}$Sb$_{12}$ is a moderate-coupling superconductor with a
superconducting gap on the entire FS.

It is apparent in Fig.~\ref{fig:rho} that the data deviate from
local BCS expression above 0.95~K. We also noticed a small hump in
$\Delta \lambda$ near 1.1~K shown in Fig.~\ref{fig:lambda}, which
shows up as a curvature change in $\rho_{s}$ in
Fig.~\ref{fig:rho}. These features may be due to inhomogeneities
and/or impurity effects in the sample, because the $T_{c}$'s of
these single crystals vary \cite{Sugawaraemail} between 1~K and
1.3~K.

It is puzzling that the substitution of Ru for Os (same column in
the periodic table) causes PrRu$_{4}$Sb$_{12}$ to differ in so
many respects from PrOs$_{4}$Sb$_{12}$. Further, evidence for gap
anisotropy in the latter compound is contradicted by $\mu$SR
measurements \cite{MacLaughlin02} (suggesting either \textit{s} or
\textit{p}-wave pairing, with $\Delta (0)$=2.1$k_{B}T_{c}$),
scanning tunneling spectroscopy \cite{Suderow03}, and NQR
measurements \cite{Kotegawa03} (with $\Delta
(0)$=2.6$k_{B}T_{c}$). If both PrRu$_{4}$Sb$_{12}$ and
PrOs$_{4}$Sb$_{12}$ have isotropic gaps, then they are unique,
especially the latter, among HFSC, suggesting the possibility of
(a) an important difference in superconducting properties between
HFSC with magnetic and non-magnetic \textit{f}-ion ground states,
and (b) a correlation between pairing symmetry (isotropic or nodal
gap) and mechanism (quadrupolar or magnetic fluctuations) of
superconductivity \cite{MacLaughlin02}.

Recently, Frederick \textit{et al.} performed X-ray powder
diffraction, magnetic susceptibility and electrical resistivity
measurements \cite{Frederick04} on single crystals of
Pr(Os$_{1-x}$Ru$_{x}$)$_{4}$Sb$_{12}$. They found that (1) the
lattice constant $a$ decreases approximately linearly with
increasing Ru concentration, (2) the splitting between the ground
and first excited state increases monotonically with $x$, with the
fits consistent with a $\Gamma_{3}$ doublet ground state for all
values of $x$, although reasonable fits can be obtained for a
$\Gamma_{1}$ ground state for $x$ near 0 and 1, and (3) $T_{c}$
decreases nearly linearly with substituent concentration away from
$x$=0 and $x$=1, but exhibits a deep minimum (0.75~K) at $x$ =
0.6. The smooth evolution of $a$ and $T_{c}$ with $x$, and the
presence of superconductivity for all values of $x$, may suggest
that both PrOs$_{4}$Sb$_{12}$ and PrRu$_{4}$Sb$_{12}$ possess the
same order-parameter symmetry. The minimum in $T_{c}$ at $x$ = 0.6
could simply mark the shift from quadrupolar-mediated heavy
fermion superconductivity to phonon-mediated BCS
superconductivity. On the other hand, one still has to contend
with measurements \cite{Izawa03,Chia03b,Aoki03,Vollmer03} that
indicate point-node gap structure, TRSB and double superconducting
transitions in PrOs$_{4}$Sb$_{12}$. If so, the minimum in $T_{c}$
could be a consequence of competing order-parameter symmetries
with a possible quantum critical point between them. It is also
interesting to notice in Ref.~\onlinecite{Frederick04} that the
step-like structure seen in ac susceptibility data in the $x$=0
sample, i.e. PrOs$_{4}$Sb$_{12}$, indicative of an intrinsic
second superconducting transition, is \textit{not} seen for all
other values of $x$.

To further elucidate the relationship between PrOs$_{4}$Sb$_{12}$
and PrRu$_{4}$Sb$_{12}$, work is underway looking at the changes
in penetration depth on Pr(Os$_{1-x}$Ru$_{x}$)$_{4}$Sb$_{12}$ for
a range of values of $x$. We want to know at which value of $x$
(paying close attention to the value $x=0.6$), if any, does the
isotropic superconducting gap evolve into a nodal one. Moreover,
we want to know whether the second superconducting transition
\cite{Vollmer03} seen in PrOs$_{4}$Sb$_{12}$ can also be seen in
samples where $0<x \le 1$. If we do not see the second
superconducting transition in the other samples, then it is
possible that the superconductivity in the two superconducting
phases of PrOs$_{4}$Sb$_{12}$ respond differently to impurities
(Ru substituents) --- one is completely destroyed by even tiny
amounts of impurities, while the other persists (though weakened)
all the way to PrRu$_{4}$Sb$_{12}$. It might be useful also if the
impurity introduced is of an element that would \textit{not}
produce an isostructural superconducting compound.

In conclusion, we report measurements of the magnetic penetration depth $%
\lambda $ in single crystals of PrRu$_{4}$Sb$_{12}$ down to 0.1~K
using a tunnel-diode based, self-inductive technique at 21 MHz.
Both $\lambda$ and $\rho_{s}$ exhibit an exponential behavior for
$T$ $<$ 0.5$T_{c}$, with parameters
$\Delta$(0)/\textit{k}$_{B}$\textit{T}$_{c}$ = 1.9 and
$\lambda(0)$ = 2900 \AA. The value of $\Delta$(0) is consistent
with the specific-heat jump value of $\Delta C/\gamma T_{c}$ =
1.87 measured elsewhere \cite{Takeda00}, while the value of
$\lambda(0)$ is consistent with the measured value \cite{Takeda00}
of the electronic heat-capacity coefficient $\gamma$, and $\Delta
(0)$. Our data are consistent with PrRu$_{4}$Sb$_{12}$ being a
moderate-coupling, fully-gapped superconductor. We also suggest
further experiments that can be done to study how the nature of
the superconducting state evolves with Ru substitution, to further
elucidate the relationship among HF behavior, superconducting gap
structure, and the presence of TRSB in the superconducting state.

This work was supported by the NSF through Grant No. DMR99-72087
and DMR01-07253, and the Grant-in Aid for Scientific Research on
the Priority Area ``Skutterudites" from MEXT in Japan. Some of the
work was carried out in the Center for Microanalysis of Materials,
University of Illinois, which is partially supported by the U.S.
Department of Energy under grant DEFG02-91-ER45439.

\bibliography{PrOs4Sb12,CeCoIn5v11}

\begin{thebibliography}{20}
\expandafter\ifx\csname natexlab\endcsname\relax\def\natexlab#1{#1}\fi
\expandafter\ifx\csname bibnamefont\endcsname\relax
  \def\bibnamefont#1{#1}\fi
\expandafter\ifx\csname bibfnamefont\endcsname\relax
  \def\bibfnamefont#1{#1}\fi
\expandafter\ifx\csname citenamefont\endcsname\relax
  \def\citenamefont#1{#1}\fi
\expandafter\ifx\csname url\endcsname\relax
  \def\url#1{\texttt{#1}}\fi
\expandafter\ifx\csname urlprefix\endcsname\relax\def\urlprefix{URL }\fi
\providecommand{\bibinfo}[2]{#2}
\providecommand{\eprint}[2][]{\url{#2}}

\bibitem[{\citenamefont{Bauer et~al.}(2002)\citenamefont{Bauer, Frederick, Ho,
  Zapf, and Maple}}]{Bauer02}
\bibinfo{author}{\bibfnamefont{E.~D.} \bibnamefont{Bauer}},
  \bibinfo{author}{\bibfnamefont{N.~A.} \bibnamefont{Frederick}},
  \bibinfo{author}{\bibfnamefont{P.-C.} \bibnamefont{Ho}},
  \bibinfo{author}{\bibfnamefont{V.~S.} \bibnamefont{Zapf}}, \bibnamefont{and}
  \bibinfo{author}{\bibfnamefont{M.~B.} \bibnamefont{Maple}},
  \bibinfo{journal}{Phys. Rev. B} \textbf{\bibinfo{volume}{65}},
  \bibinfo{pages}{100506 (R)} (\bibinfo{year}{2002}).

\bibitem[{\citenamefont{Maple et~al.}(2002)\citenamefont{Maple, Ho, Zapf,
  Frederick, Bauer, Yuhasz, Woodward, and Lynn}}]{Maple02}
\bibinfo{author}{\bibfnamefont{M.~B.} \bibnamefont{Maple}},
  \bibinfo{author}{\bibfnamefont{P.-C.} \bibnamefont{Ho}},
  \bibinfo{author}{\bibfnamefont{V.~S.} \bibnamefont{Zapf}},
  \bibinfo{author}{\bibfnamefont{N.~A.} \bibnamefont{Frederick}},
  \bibinfo{author}{\bibfnamefont{E.~D.} \bibnamefont{Bauer}},
  \bibinfo{author}{\bibfnamefont{W.~M.} \bibnamefont{Yuhasz}},
  \bibinfo{author}{\bibfnamefont{F.~M.} \bibnamefont{Woodward}},
  \bibnamefont{and} \bibinfo{author}{\bibfnamefont{J.~W.} \bibnamefont{Lynn}},
  \bibinfo{journal}{J. Phys. Soc. Jpn., Suppl. B}
  \textbf{\bibinfo{volume}{71}}, \bibinfo{pages}{23} (\bibinfo{year}{2002}).

\bibitem[{\citenamefont{Takeda and Ishikawa}(2000)}]{Takeda00}
\bibinfo{author}{\bibfnamefont{N.}~\bibnamefont{Takeda}} \bibnamefont{and}
  \bibinfo{author}{\bibfnamefont{M.}~\bibnamefont{Ishikawa}},
  \bibinfo{journal}{J. Phys. Soc. Jpn.} \textbf{\bibinfo{volume}{69}},
  \bibinfo{pages}{868} (\bibinfo{year}{2000}).

\bibitem[{\citenamefont{Kotegawa et~al.}(2003)\citenamefont{Kotegawa, Yogi,
  Imamura, Kawasaki, q.~Zheng, Kitaoka, Ohsaki, Sugawara, Aoki, and
  Sato}}]{Kotegawa03}
\bibinfo{author}{\bibfnamefont{H.}~\bibnamefont{Kotegawa}},
  \bibinfo{author}{\bibfnamefont{M.}~\bibnamefont{Yogi}},
  \bibinfo{author}{\bibfnamefont{Y.}~\bibnamefont{Imamura}},
  \bibinfo{author}{\bibfnamefont{Y.}~\bibnamefont{Kawasaki}},
  \bibinfo{author}{\bibfnamefont{G.}~\bibnamefont{q.~Zheng}},
  \bibinfo{author}{\bibfnamefont{Y.}~\bibnamefont{Kitaoka}},
  \bibinfo{author}{\bibfnamefont{S.}~\bibnamefont{Ohsaki}},
  \bibinfo{author}{\bibfnamefont{H.}~\bibnamefont{Sugawara}},
  \bibinfo{author}{\bibfnamefont{Y.}~\bibnamefont{Aoki}}, \bibnamefont{and}
  \bibinfo{author}{\bibfnamefont{H.}~\bibnamefont{Sato}},
  \bibinfo{journal}{Phys. Rev. Lett.} \textbf{\bibinfo{volume}{90}},
  \bibinfo{pages}{027001} (\bibinfo{year}{2003}).

\bibitem[{\citenamefont{Yogi et~al.}(2003)\citenamefont{Yogi, Kotegawa,
  Imamura, q.~Zheng, Kitaoka, Sugawara, and Sato}}]{Yogi03}
\bibinfo{author}{\bibfnamefont{M.}~\bibnamefont{Yogi}},
  \bibinfo{author}{\bibfnamefont{H.}~\bibnamefont{Kotegawa}},
  \bibinfo{author}{\bibfnamefont{Y.}~\bibnamefont{Imamura}},
  \bibinfo{author}{\bibfnamefont{G.}~\bibnamefont{q.~Zheng}},
  \bibinfo{author}{\bibfnamefont{Y.}~\bibnamefont{Kitaoka}},
  \bibinfo{author}{\bibfnamefont{H.}~\bibnamefont{Sugawara}}, \bibnamefont{and}
  \bibinfo{author}{\bibfnamefont{H.}~\bibnamefont{Sato}},
  \bibinfo{journal}{Phys. Rev. B} \textbf{\bibinfo{volume}{67}},
  \bibinfo{pages}{180501(R)} (\bibinfo{year}{2003}).

\bibitem[{\citenamefont{Chia et~al.}(2003{\natexlab{a}})\citenamefont{Chia,
  Salamon, Sugawara, and Sato}}]{Chia03b}
\bibinfo{author}{\bibfnamefont{E.~E.~M.} \bibnamefont{Chia}},
  \bibinfo{author}{\bibfnamefont{M.~B.} \bibnamefont{Salamon}},
  \bibinfo{author}{\bibfnamefont{H.}~\bibnamefont{Sugawara}}, \bibnamefont{and}
  \bibinfo{author}{\bibfnamefont{H.}~\bibnamefont{Sato}},
  \bibinfo{journal}{Phys. Rev. Lett.} \textbf{\bibinfo{volume}{91}},
  \bibinfo{pages}{247003} (\bibinfo{year}{2003}{\natexlab{a}}).

\bibitem[{\citenamefont{Izawa et~al.}(2003)\citenamefont{Izawa, Nakajima,
  Goryo, Matsuda, Osaki, Sugawara, Sato, Thalmeier, and Maki}}]{Izawa03}
\bibinfo{author}{\bibfnamefont{K.}~\bibnamefont{Izawa}},
  \bibinfo{author}{\bibfnamefont{Y.}~\bibnamefont{Nakajima}},
  \bibinfo{author}{\bibfnamefont{J.}~\bibnamefont{Goryo}},
  \bibinfo{author}{\bibfnamefont{Y.}~\bibnamefont{Matsuda}},
  \bibinfo{author}{\bibfnamefont{S.}~\bibnamefont{Osaki}},
  \bibinfo{author}{\bibfnamefont{H.}~\bibnamefont{Sugawara}},
  \bibinfo{author}{\bibfnamefont{H.}~\bibnamefont{Sato}},
  \bibinfo{author}{\bibfnamefont{P.}~\bibnamefont{Thalmeier}},
  \bibnamefont{and} \bibinfo{author}{\bibfnamefont{K.}~\bibnamefont{Maki}},
  \bibinfo{journal}{Phys. Rev. Lett.} \textbf{\bibinfo{volume}{90}},
  \bibinfo{pages}{117001} (\bibinfo{year}{2003}).

\bibitem[{\citenamefont{Aoki et~al.}(2003)\citenamefont{Aoki, Tsuchiya,
  Kanayama, Saha, Sugawara, Sato, Higemoto, Koda, Ohishi, Nishiyama
  et~al.}}]{Aoki03}
\bibinfo{author}{\bibfnamefont{Y.}~\bibnamefont{Aoki}},
  \bibinfo{author}{\bibfnamefont{A.}~\bibnamefont{Tsuchiya}},
  \bibinfo{author}{\bibfnamefont{T.}~\bibnamefont{Kanayama}},
  \bibinfo{author}{\bibfnamefont{S.~R.} \bibnamefont{Saha}},
  \bibinfo{author}{\bibfnamefont{H.}~\bibnamefont{Sugawara}},
  \bibinfo{author}{\bibfnamefont{H.}~\bibnamefont{Sato}},
  \bibinfo{author}{\bibfnamefont{W.}~\bibnamefont{Higemoto}},
  \bibinfo{author}{\bibfnamefont{A.}~\bibnamefont{Koda}},
  \bibinfo{author}{\bibfnamefont{K.}~\bibnamefont{Ohishi}},
  \bibinfo{author}{\bibfnamefont{K.}~\bibnamefont{Nishiyama}},
  \bibnamefont{et~al.}, \bibinfo{journal}{Phys. Rev. Lett.}
  \textbf{\bibinfo{volume}{91}}, \bibinfo{pages}{067003}
  (\bibinfo{year}{2003}).

\bibitem[{\citenamefont{Matsuda et~al.}(2002)\citenamefont{Matsuda, Abe,
  Watanuki, Sugawara, Aoki, Sato, Inada, Settai, and
  $\bar{O}$nuki}}]{Matsuda02}
\bibinfo{author}{\bibfnamefont{T.~D.} \bibnamefont{Matsuda}},
  \bibinfo{author}{\bibfnamefont{K.}~\bibnamefont{Abe}},
  \bibinfo{author}{\bibfnamefont{F.}~\bibnamefont{Watanuki}},
  \bibinfo{author}{\bibfnamefont{H.}~\bibnamefont{Sugawara}},
  \bibinfo{author}{\bibfnamefont{Y.}~\bibnamefont{Aoki}},
  \bibinfo{author}{\bibfnamefont{H.}~\bibnamefont{Sato}},
  \bibinfo{author}{\bibfnamefont{Y.}~\bibnamefont{Inada}},
  \bibinfo{author}{\bibfnamefont{R.}~\bibnamefont{Settai}}, \bibnamefont{and}
  \bibinfo{author}{\bibfnamefont{Y.}~\bibnamefont{$\bar{O}$nuki}},
  \bibinfo{journal}{Physica B} \textbf{\bibinfo{volume}{312-313}},
  \bibinfo{pages}{832} (\bibinfo{year}{2002}).

\bibitem[{\citenamefont{Bonalde et~al.}(2000)\citenamefont{Bonalde, Yanoff,
  Salamon, Harlingen, Chia, Mao, and Maeno}}]{Bonalde2000}
\bibinfo{author}{\bibfnamefont{I.}~\bibnamefont{Bonalde}},
  \bibinfo{author}{\bibfnamefont{B.~D.} \bibnamefont{Yanoff}},
  \bibinfo{author}{\bibfnamefont{M.~B.} \bibnamefont{Salamon}},
  \bibinfo{author}{\bibfnamefont{D.~J.~V.} \bibnamefont{Harlingen}},
  \bibinfo{author}{\bibfnamefont{E.~M.~E.} \bibnamefont{Chia}},
  \bibinfo{author}{\bibfnamefont{Z.~Q.} \bibnamefont{Mao}}, \bibnamefont{and}
  \bibinfo{author}{\bibfnamefont{Y.}~\bibnamefont{Maeno}},
  \bibinfo{journal}{Phys. Rev. Lett.} \textbf{\bibinfo{volume}{85}},
  \bibinfo{pages}{4775} (\bibinfo{year}{2000}).

\bibitem[{\citenamefont{Chia et~al.}(2003{\natexlab{b}})\citenamefont{Chia,
  Harlingen, Salamon, Yanoff, Bonalde, and Sarrao}}]{Chia03}
\bibinfo{author}{\bibfnamefont{E.~E.~M.} \bibnamefont{Chia}},
  \bibinfo{author}{\bibfnamefont{D.~J.~V.} \bibnamefont{Harlingen}},
  \bibinfo{author}{\bibfnamefont{M.~B.} \bibnamefont{Salamon}},
  \bibinfo{author}{\bibfnamefont{B.~D.} \bibnamefont{Yanoff}},
  \bibinfo{author}{\bibfnamefont{I.}~\bibnamefont{Bonalde}}, \bibnamefont{and}
  \bibinfo{author}{\bibfnamefont{J.~L.} \bibnamefont{Sarrao}},
  \bibinfo{journal}{Phys. Rev. B} \textbf{\bibinfo{volume}{67}},
  \bibinfo{pages}{014527} (\bibinfo{year}{2003}{\natexlab{b}}).

\bibitem[{\citenamefont{Prozorov et~al.}(2000)\citenamefont{Prozorov, Gianetta,
  Carrington, and Araujo-Moreira}}]{Prozorov2000}
\bibinfo{author}{\bibfnamefont{R.}~\bibnamefont{Prozorov}},
  \bibinfo{author}{\bibfnamefont{R.~W.} \bibnamefont{Gianetta}},
  \bibinfo{author}{\bibfnamefont{A.}~\bibnamefont{Carrington}},
  \bibnamefont{and} \bibinfo{author}{\bibfnamefont{F.~M.}
  \bibnamefont{Araujo-Moreira}}, \bibinfo{journal}{Phys. Rev. B}
  \textbf{\bibinfo{volume}{62}}, \bibinfo{pages}{115} (\bibinfo{year}{2000}).

\bibitem[{\citenamefont{Orlando et~al.}(1979)\citenamefont{Orlando,
  E.~J.~McNiff, Foner, and Beasley}}]{Orlando1979}
\bibinfo{author}{\bibfnamefont{T.~P.} \bibnamefont{Orlando}},
  \bibinfo{author}{\bibfnamefont{J.}~\bibnamefont{E.~J.~McNiff}},
  \bibinfo{author}{\bibfnamefont{S.}~\bibnamefont{Foner}}, \bibnamefont{and}
  \bibinfo{author}{\bibfnamefont{M.~R.} \bibnamefont{Beasley}},
  \bibinfo{journal}{Phys. Rev. B} \textbf{\bibinfo{volume}{19}},
  \bibinfo{pages}{4545} (\bibinfo{year}{1979}).

\bibitem[{\citenamefont{Kresin and Parkhomenko}(1975)}]{Kresin1975}
\bibinfo{author}{\bibfnamefont{V.~Z.} \bibnamefont{Kresin}} \bibnamefont{and}
  \bibinfo{author}{\bibfnamefont{V.~P.} \bibnamefont{Parkhomenko}},
  \bibinfo{journal}{Sov. Phys.- Solid State} \textbf{\bibinfo{volume}{16}},
  \bibinfo{pages}{2180} (\bibinfo{year}{1975}).

\bibitem[{\citenamefont{Gross et~al.}(1986)\citenamefont{Gross, Chandrasekhar,
  Einzel, Andres, Hirschfeld, Ott, Beuers, Fisk, and Smith}}]{Gross1986}
\bibinfo{author}{\bibfnamefont{F.}~\bibnamefont{Gross}},
  \bibinfo{author}{\bibfnamefont{B.~S.} \bibnamefont{Chandrasekhar}},
  \bibinfo{author}{\bibfnamefont{D.}~\bibnamefont{Einzel}},
  \bibinfo{author}{\bibfnamefont{K.}~\bibnamefont{Andres}},
  \bibinfo{author}{\bibfnamefont{P.~J.} \bibnamefont{Hirschfeld}},
  \bibinfo{author}{\bibfnamefont{H.~R.} \bibnamefont{Ott}},
  \bibinfo{author}{\bibfnamefont{J.}~\bibnamefont{Beuers}},
  \bibinfo{author}{\bibfnamefont{Z.}~\bibnamefont{Fisk}}, \bibnamefont{and}
  \bibinfo{author}{\bibfnamefont{J.~L.} \bibnamefont{Smith}},
  \bibinfo{journal}{Z. Phys. B} \textbf{\bibinfo{volume}{64}},
  \bibinfo{pages}{175} (\bibinfo{year}{1986}).

\bibitem[{\citenamefont{Sugawara}(2003)}]{Sugawaraemail}
\bibinfo{author}{\bibfnamefont{H.}~\bibnamefont{Sugawara}},
  \bibinfo{journal}{private conversation}  (\bibinfo{year}{2003}).

\bibitem[{\citenamefont{MacLaughlin et~al.}(2002)\citenamefont{MacLaughlin,
  Sonier, Heffner, Bernal, Young, Rose, Morris, Bauer, Do, and
  Maple}}]{MacLaughlin02}
\bibinfo{author}{\bibfnamefont{D.~E.} \bibnamefont{MacLaughlin}},
  \bibinfo{author}{\bibfnamefont{J.~E.} \bibnamefont{Sonier}},
  \bibinfo{author}{\bibfnamefont{R.~H.} \bibnamefont{Heffner}},
  \bibinfo{author}{\bibfnamefont{O.~O.} \bibnamefont{Bernal}},
  \bibinfo{author}{\bibfnamefont{B.-L.} \bibnamefont{Young}},
  \bibinfo{author}{\bibfnamefont{M.~S.} \bibnamefont{Rose}},
  \bibinfo{author}{\bibfnamefont{G.~D.} \bibnamefont{Morris}},
  \bibinfo{author}{\bibfnamefont{E.~D.} \bibnamefont{Bauer}},
  \bibinfo{author}{\bibfnamefont{T.~D.} \bibnamefont{Do}}, \bibnamefont{and}
  \bibinfo{author}{\bibfnamefont{M.~B.} \bibnamefont{Maple}},
  \bibinfo{journal}{Phys. Rev. Lett.} \textbf{\bibinfo{volume}{89}},
  \bibinfo{pages}{157001} (\bibinfo{year}{2002}).

\bibitem[{\citenamefont{Suderow et~al.}(2003)\citenamefont{Suderow, Vieira,
  Strand, Bud'ko, and Canfield}}]{Suderow03}
\bibinfo{author}{\bibfnamefont{H.}~\bibnamefont{Suderow}},
  \bibinfo{author}{\bibfnamefont{S.}~\bibnamefont{Vieira}},
  \bibinfo{author}{\bibfnamefont{J.~D.} \bibnamefont{Strand}},
  \bibinfo{author}{\bibfnamefont{S.}~\bibnamefont{Bud'ko}}, \bibnamefont{and}
  \bibinfo{author}{\bibfnamefont{P.~C.} \bibnamefont{Canfield}},
  \bibinfo{journal}{cond-mat/0306463}  (\bibinfo{year}{2003}).

\bibitem[{\citenamefont{Frederick et~al.}(2004)\citenamefont{Frederick, Do, Ho,
  Butch, Zapf, and Maple}}]{Frederick04}
\bibinfo{author}{\bibfnamefont{N.~A.} \bibnamefont{Frederick}},
  \bibinfo{author}{\bibfnamefont{T.~D.} \bibnamefont{Do}},
  \bibinfo{author}{\bibfnamefont{P.-C.} \bibnamefont{Ho}},
  \bibinfo{author}{\bibfnamefont{N.~P.} \bibnamefont{Butch}},
  \bibinfo{author}{\bibfnamefont{V.~S.} \bibnamefont{Zapf}}, \bibnamefont{and}
  \bibinfo{author}{\bibfnamefont{M.~B.} \bibnamefont{Maple}},
  \bibinfo{journal}{Phys. Rev. B} \textbf{\bibinfo{volume}{69}},
  \bibinfo{pages}{024523} (\bibinfo{year}{2004}).

\bibitem[{\citenamefont{Vollmer et~al.}(2003)\citenamefont{Vollmer,
  Fai$\beta$t, Pfleiderer, v.~Lohneysen, Bauer, Ho, Zapf, and
  Maple}}]{Vollmer03}
\bibinfo{author}{\bibfnamefont{R.}~\bibnamefont{Vollmer}},
  \bibinfo{author}{\bibfnamefont{A.}~\bibnamefont{Fai$\beta$t}},
  \bibinfo{author}{\bibfnamefont{C.}~\bibnamefont{Pfleiderer}},
  \bibinfo{author}{\bibfnamefont{H.}~\bibnamefont{v.~Lohneysen}},
  \bibinfo{author}{\bibfnamefont{E.~D.} \bibnamefont{Bauer}},
  \bibinfo{author}{\bibfnamefont{P.-C.} \bibnamefont{Ho}},
  \bibinfo{author}{\bibfnamefont{V.}~\bibnamefont{Zapf}}, \bibnamefont{and}
  \bibinfo{author}{\bibfnamefont{M.~B.} \bibnamefont{Maple}},
  \bibinfo{journal}{Phys. Rev. Lett.} \textbf{\bibinfo{volume}{90}},
  \bibinfo{pages}{057001} (\bibinfo{year}{2003}).

\end{thebibliography}
\bigskip

\end{document}